\documentclass[sigconf]{acmart}

\usepackage{framed}
\usepackage[strict]{changepage}
\usepackage{subcaption}
\usepackage{placeins}
\usepackage{lettrine} 
\definecolor{formalshade}{rgb}{0.95,0.95,1}
\definecolor{darkblue}{rgb}{0.36,0.54,0.66}

\AtBeginDocument{%
  }

\copyrightyear{2026}
\acmYear{2026}
\setcopyright{cc}
\setcctype{by-nc-nd}
\acmConference[CHI EA '26]{Extended Abstracts of the 2026 CHI Conference on Human Factors in Computing Systems}{April 13--17, 2026}{Barcelona, Spain}
\acmBooktitle{Extended Abstracts of the 2026 CHI Conference on Human Factors in Computing Systems (CHI EA '26), April 13--17, 2026, Barcelona, Spain}
\acmDOI{10.1145/3772363.3799265}
\acmISBN{979-8-4007-2281-3/2026/04}

\usepackage{subcaption}
\usepackage{placeins}
\begin{document}

\title[Beyond Faders]{Beyond Faders: Understanding 6DoF Gesture \\Ecologies in Music Mixing}


\author{Jeremy Wertheim C. Chen}
\orcid{0009-0007-3660-9327}
\affiliation{%
  \institution{De La Salle University - Manila}
  \city{Manila}
  \country{Philippines}
}
\email{jeremy_chen@dlsu.edu.ph}

\author{Rendell Christian J. Ngo}
\orcid{0009-0001-8239-2233}
\affiliation{%
 \institution{De La Salle University - Manila}
  \city{Manila}
  \country{Philippines}
}
\email{rendell_christian_ngo@dlsu.edu.ph}

\author{Cedric Matthew B. Yu}
\orcid{0009-0009-0000-0055}
\affiliation{%
  \institution{De La Salle University - Manila}
  \city{Manila}
  \country{Philippines}
}
\email{ced_yu@dlsu.edu.ph}

\author{Hans Emilio M. Lumagui}
\orcid{0009-0001-0340-4655}
\affiliation{%
  \institution{De La Salle University - Manila}
  \city{Manila}
  \country{Philippines}
}
\email{hans_lumagui@dlsu.edu.ph}

\author{Ethan R. Badayos}
\orcid{0009-0008-7725-0792}
\affiliation{%
  \institution{De La Salle University - Manila}
  \city{Manila}
  \country{Philippines}
}
\email{ethan_r_badayos@dlsu.edu.ph}

\author{Jordan Aiko Deja} 
\orcid{0000-0001-9341-6088}
 \affiliation{%
  \institution{De La Salle University}
  \city{Manila}
  \country{Philippines}
  }
\email{jordan.deja@dlsu.edu.ph}


\begin{abstract}

Extended reality (XR) enables new music-mixing workflows by moving beyond 2D faders toward embodied, spatial interaction. However, it remains unclear which six-degree-of-freedom (6DoF) gestures align with real-world mixing practices and whether such interactions support manageable cognitive load and positive user experience. We conducted a design workshop with experienced mixers to elicit gesture concepts for core audio tasks gain, compression, equalization, and automation, and implemented these in an XR prototype. A user study (n=12) evaluated the ecological validity of the gestures using cognitive load measures, user-experience ratings, and interviews. Participants generally found 6DoF gestures intuitive and well-mapped to mixing tasks, reporting strong immersion and a sense of connection with the audio environment. Cognitive load differences across gestures were minimal, though participants expressed preferences shaped by workflow familiarity and perceived control. We discuss implications for designing XR mixing tools that balance expressiveness, precision, and ecological validity.
\end{abstract}


\begin{teaserfigure}
  \includegraphics[width=\textwidth]{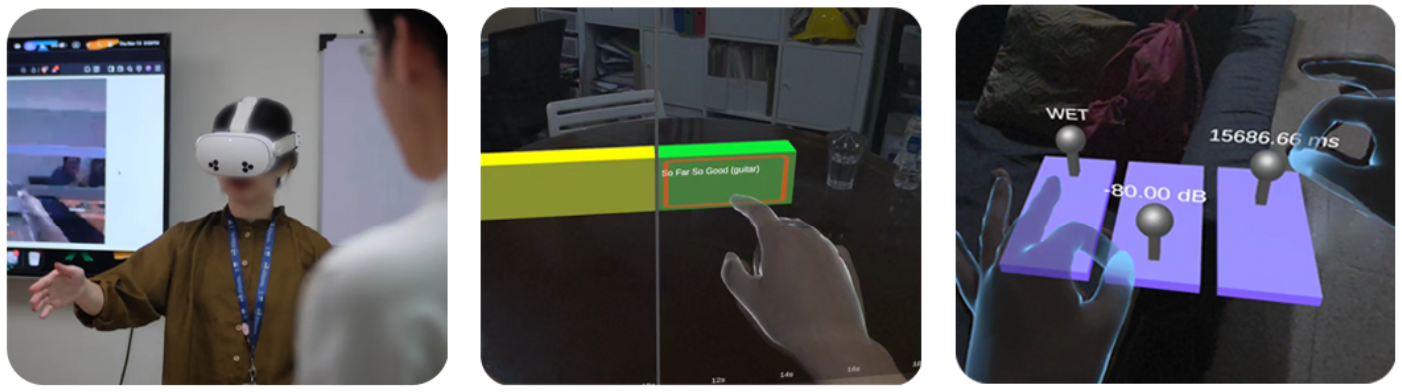}
  \caption{\textbf{Ecological study of 6DoF gesture interaction in music mixing.} Left: In-situ interaction with the XR mixing prototype. Middle: Timeline-based parameter visualization used during mixing tasks. Right: Spatial levers enabling 6DoF control of audio parameters.}
  \label{fig:teaser}
\end{teaserfigure}



\renewcommand{\shortauthors}{Jeremy Weirtheim Chan et al.}
\maketitle

\section{Introduction and Background}
\label{sec:intro}

\par Digital Audio Workstations (DAWs) are the dominant tools for contemporary music mixing, providing flexible and powerful control over audio production \cite{loveridge2023overview}. However, DAW interfaces are largely constrained to two-dimensional (2D) screen-based interactions, typically mediated through a mouse or keyboard. While effective, these interaction paradigms limit embodied engagement and expressive control, particularly when compared to the tactile manipulation afforded by physical mixers and analog consoles \cite{gelineck2013towards-a,RiddershomBargum2023}. As a result, mixers often experience a separation between bodily action and auditory outcome, which can affect immersion and creative flow. For instance, several mixers report that as they usually do their mixing, they get in the flow and feel like dancing with the groove. As they are manipulating and moving between 2D sliders, movement is often limited and constrain within these dimensions. We argue that by using 6DoF which are usually present in 3D environments, we can expand these interactions and in-the-groove sensations. As seen in prior studies, users have to be really comfortable with the interface before they actually reap the benefits of using the interface~\cite{kim2012designing}. We argue that extending these from 2D to 3D in music mixing can be explored too. 

\par Extended Reality (XR) systems offer opportunities to reimagine music-mixing workflows through immersive, spatial interaction. Prior work has shown that virtual environments can enhance musical engagement by allowing users to interact with sound sources as spatial objects rather than abstract interface elements \cite{gugenheimer2017sharevr,lin2017design,safikhani2022immersive}. In the context of music production, VR-based systems have demonstrated benefits for immersion, spatial reasoning, and creative exploration \cite{gospodarek2024acoustic,loveridge2023overview}. For example, \citeN{RiddershomBargum2023} explored spatial sound manipulation in VR, showing how moving sound sources in 3D space can offer more intuitive control than traditional panning sliders.

\par XR interaction is further enabled by six degrees of freedom (6DoF), combining three-dimensional translation with rotation around each axis \cite{mrozProduction2021}. Prior research suggests that 6DoF interaction can support intuitive learning, expressive control, and creativity in musical and sound design contexts \cite{llewellyn2021towards,ppali2022keep,li2022effect}. Systems such as AudioMiXR demonstrate how spatial manipulation of audio objects using 6DoF can support immersive sound design and positive user experience \cite{woodard2025audiomixr}. However, existing studies primarily focus on exploratory sound design or performance, rather than structured, real-world mixing workflows.

\par In parallel, gesture-based interaction has been widely studied in HCI as a natural and embodied input modality \cite{vuletic2019systematic,holdengreber2021intuitive}. Prior systems such as Siftables \cite{merrill2007siftables}, Skinput \cite{harrison2010skinput}, and Project Digits \cite{kim2012digits} demonstrate how physical movement and rotation can be directly mapped to digital control. In XR, hand tracking has become a common interaction technique, enabling direct manipulation of virtual objects without handheld controllers \cite{voigt2020influence,meta2025handtracking}. While gesture-based input has been applied to creative tasks such as drawing, performance, and audio manipulation \cite{kim2013ar,rosa2016fast,castro2023exploring}, challenges remain in ensuring precision, discoverability, and appropriate gesture–function mappings \cite{vuletic2019systematic}.

\par Despite growing interest in XR-based music tools, there remains limited empirical evidence on how 6DoF gestures align with established music-mixing practices. In this work, we operationalize ecological validity as the degree to which spatial gestures align with users’ expectations of mixing workflows, support embodied comfort and control, and are perceived as appropriate for specific audio parameters. In particular, it is unclear which spatial gestures are perceived as intuitive and ecologically valid when applied to core mixing tasks such as gain, compression, equalization, and automation. Prior work highlights the potential of spatial interaction for creativity and immersion, but stops short of systematically evaluating gesture suitability across specific mixing parameters.

\par Building on earlier workshop-based design exploration \cite{sonicmixr2025}, this study evaluates how selected 6DoF gestures support user experience in music-mixing tasks. We investigate the following research question: \textbf{How well do specific 6DoF gestures align with user experience for controlling different music mixing elements?}

\par
\begin{figure*}[t]
    \centering
    \includegraphics[width=\linewidth]{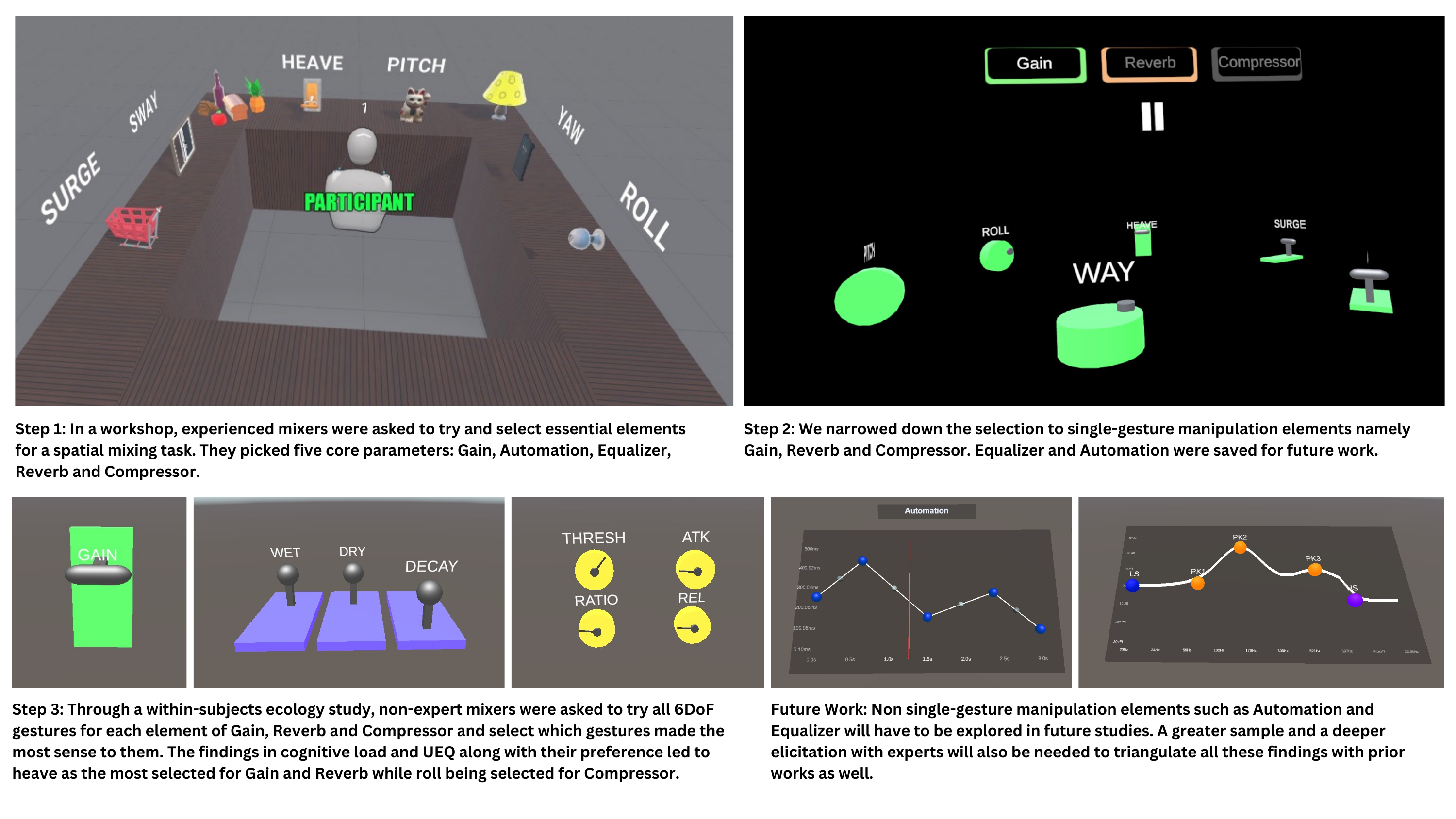}
\caption{\textbf{Three-stage evaluation of 6DoF gesture mappings for music mixing.} (1) Workshop elicitation of core mixing elements. (2) Selection of single-gesture parameters for XR implementation. (3) Within-subject ecology study assessing gesture preference, cognitive load, and user experience. Heave was most preferred for Gain and Reverb, and roll for Compressor.}
    \label{fig:data_all}
\end{figure*}

 To address this question, we integrate workshop-elicited gestures into an XR prototype and conduct a controlled user study evaluating cognitive load, user experience, and qualitative perceptions. Through this work, we aim to contribute empirical insights into the ecological validity of 6DoF gesture interaction for music mixing and inform the design of future XR-based audio tools. Figure~\ref{fig:data_all} summarizes our contribution into a visual narrative of all the steps and core findings from this paper.

\section{Validating 6DoF in Music Mixing}

\subsection{Design Workshop}
\par This study builds on a prior design workshop reported by \citeN{sonicmixr2025}, from which we derive gesture concepts and sound elements relevant to music mixing in XR. Here, we summarize only the workshop outcomes necessary to motivate and ground the present evaluation.

\par The workshop was structured into three phases: \textbf{Processes}, \textbf{Tools}, and \textbf{Conceptualization}. Our work focuses on findings from the latter two phases, where participants identified key mixing elements and proposed gesture-based interaction strategies for XR environments. Participants consisted of three music mixers with varying levels of expertise, including casual mixers (less than one year of experience or hobbyist practice) and expert mixers with professional or semi-professional experience. Participants collaboratively identified essential elements for a spatial mixing environment based on prior literature and their own workflow descriptions. The final set of elements was constrained to a maximum of eight and resulted in five core mixing parameters: \textbf{Gain}, \textbf{Automation}, \textbf{Equalizer}, \textbf{Reverb}, and \textbf{Compressor}.

\par To support gesture elicitation, a low-fidelity mock-up created in ShapesXR was used to represent six manipulable objects corresponding to translational and rotational axes in 6DoF. Participants were asked to imagine controlling each mixing element through spatial manipulation of these objects. For each mixing element, participants selected their top two preferred 6DoF axes based on perceived intuitiveness and expressive fit. First choices were assigned two points and second choices one point. In cases of tied scores, a coin flip was used to determine the selected gesture. The design workshop yielded a broader set of candidate mixing elements and gesture mappings than could be feasibly evaluated within a controlled study. To ensure focused and comparable evaluation, we narrowed the scope to a subset of elements that are both central to mixing practice and amenable to single-gesture manipulation in XR.

\par Based on the workshop outcomes, we selected \textit{Gain}, \textit{Reverb}, and \textit{Compressor} for implementation and evaluation. These elements represent commonly used, continuous parameters in music mixing and can be meaningfully controlled using isolated 6DoF gestures. More complex elements such as \textit{Equalization} and \textit{Automation} were excluded, as they typically require multi-point or compound interactions that fall outside the scope of single-gesture evaluation in this study.



\subsection{Ecological Validation Study}
\par We conducted a within-subjects user study with 12 participants recruited through convenience sampling. Participants were non-mixers, defined as individuals with no prior experience in music production, audio mixing, or use of digital audio workstations or mixing consoles.

\par Participants wore a VR headset and interacted with the XR mixing tool using hand-tracked gestures, while standing up. For each mixing element, participants explored the available gesture options and were asked to identify their two preferred gestures based on perceived intuitiveness and suitability for controlling the element. Preferences were elicited using the prompt: \textit{``For controlling [element], which gesture makes the most sense to you?''}

\par After completing interaction with all gesture options for a given element, participants completed the NASA-TLX to assess cognitive load, followed by the UEQ-S to capture user experience. Participants then assigned two points to their most preferred gesture and one point to their second choice. This procedure was repeated for \textit{Gain}, \textit{Reverb}, and \textit{Compressor}.

\par Upon completing all tasks, participants took part in a semi-structured interview focused on their gesture preferences, perceived control, and reasoning behind their selections. Interviews were used to contextualize quantitative findings and identify patterns related to workflow familiarity and gesture–task alignment. Participants were randomly assigned to each sequence. Assignment was performed using a draw-from-box procedure without replacement to ensure balanced exposure across conditions.

\section{Findings}

\par \textit{Gesture Preference.}
Gesture preferences were analyzed for 12 non-mixer participants across \textit{Gain}, \textit{Reverb}, and \textit{Compressor} using a point-allocation scheme. Across all three elements, \textit{heave} received the highest total preference score among non-mixers. Fisher’s exact tests comparing these preferences with mixer data from the prior workshop \cite{sonicmixr2025} but were not significant for \textit{Gain} ($p > 0.05$), but significant differences for \textit{Reverb} and \textit{Compressor} ($p < 0.05$).

\par Standardized residuals indicate where preferences diverged between groups. For \textit{Reverb}, non-mixers strongly preferred \textit{heave} (2.79), whereas mixers avoided \textit{heave} ($-2.79$) and favored \textit{surge} (2.00) and \textit{pitch} (2.32). For \textit{Compressor}, mixers showed strong preferences for \textit{roll} (2.57) and \textit{pitch} (2.01).

\par \textit{Cognitive Load.}
Cognitive load was measured after each element using NASA-TLX. A repeated-measures ANOVA revealed were not significant in perceived workload across \textit{Gain}, \textit{Reverb}, and \textit{Compressor} ($p > 0.05$).

\par \textit{User Experience.}
User experience was assessed using the UEQ-S. \textit{Gain} received lower ratings (PQ = 1.083, HQ = 1.197, Overall = 1.500; “Good”) compared to \textit{Reverb} and \textit{Compressor}, both of which achieved “Excellent” benchmark ratings \cite{schrepp2017construction}. Friedman tests showed a significant difference in overall experience across elements ($p < 0.05$), with post-hoc comparisons indicating a significant difference between \textit{Gain} and \textit{Compressor}. Pragmatic quality also differed significantly ($p < 0.05$), with \textit{Gain} rated as less usable than both \textit{Reverb} and \textit{Compressor}.

\section{Discussion and Design Implications}

\par Our findings indicate that gesture preferences in XR music mixing are shaped by prior mixing experience, highlighting an important dimension of ecological validity. While both mixers and non-mixers preferred \textit{heave} for controlling \textit{Gain}, preferences diverged for \textit{Reverb} and \textit{Compressor}. Non-mixers continued to favor \textit{heave}, whereas mixers preferred \textit{surge} and \textit{roll}, respectively.

\par Interview data help explain these differences. Non-mixers frequently framed gesture choices around comfort and immediate intuitiveness, describing rotational gestures as awkward or tiring. In contrast, mixers appeared more tolerant of less comfortable gestures when these aligned with familiar studio metaphors and learned interaction patterns. As one mixer explained, \textit{“I pick surge first because I imagine a mixer on the table.”} This distinction echoes prior observations that expert users often rely on internalized domain conventions rather than purely embodied cues when interacting with new interfaces \cite{jost2019quantitative,atsikpasi2022scoping}.

\par \textbf{Insight 1: Ecological validity is expertise-dependent.}
Gestures that align with professional workflows may not be perceived as intuitive or comfortable by non-experts. Designing XR audio tools therefore involves balancing domain-specific interaction metaphors with gestures that remain learnable and physically accessible to novice users, a tension also observed in prior studies of expertise-sensitive interaction design \cite{atsikpasi2022scoping,jost2019quantitative}.

\par Across elements, participants reported generally positive engagement, with higher user experience scores for \textit{Reverb} and \textit{Compressor} than for \textit{Gain}. Interview responses suggest that engagement was driven by immediate auditory feedback and a sense of direct control over sound, supporting prior findings that tight perception–action coupling is central to immersive and expressive interaction.

\par At the same time, comfort emerged as a limiting factor for expressiveness. Rotational gestures were often associated with fatigue and reduced ease of use, occasionally disrupting immersion, whereas translational gestures were consistently described as natural and effortless. This pattern aligns with early human–computer interaction work showing that increasing control dimensionality can introduce coordination and physical costs \cite{buxton1983lexical,card1991morphological}.

\par \textbf{Insight 2: More degrees of freedom do not necessarily increase expressiveness.}
Although 6DoF interaction expands the interaction space, additional degrees of freedom—particularly rotational ones—may introduce physical strain or cognitive overhead that undermines perceived expressiveness and immersion. Recent work on mid-air and gestural interaction similarly reports tradeoffs between expressiveness and physical effort \cite{javerliat2024plume}. Given that our validity study was done with a small sample of participants, determining the ``right'' gestures for a specific element manipulation might warrant further exploration. 

\par Participants also described a short learning curve, reporting greater ease of use as they became familiar with the gesture mappings. Understanding gesture–sound relationships emerged through repeated exploration rather than explicit instruction, with participants describing learning through trying different motions and listening to their effects.

\par \textbf{Insight 3: Gesture meaning emerges through use, not explanation.}
Rather than relying solely on predefined metaphors or tutorials, XR mixing systems may benefit from supporting exploratory interaction that allows users to develop embodied understanding through action and feedback, consistent with prior work on designing for embodied learning and discovery \cite{ismail2018designing}.

\par Together, these insights suggest that XR music-mixing systems should support adaptable gesture sets, prioritize physical comfort as a first-class design constraint, and account for differences in expertise when mapping gestures to audio parameters.

\section{Limitations and Future Work}

\par This study comes with several limitations. First, minor inconsistencies in gesture responsiveness and latency may have influenced participants’ perceptions of certain gestures. Improving interaction fidelity could help better isolate gesture effects. Second, the study was conducted in a room setting while standing up, rather than a real-world studio environment with all applicable surrounding equipment. Future evaluations situated in authentic mixing contexts may better capture ecological validity and long-term use. Finally, our participant pool consisted of 12 non-mixers, which limits generalizability across levels of expertise. Future work should include professional mixers and larger samples to examine how gesture preferences differ with experience and evolve over time. Our paper presents an initial groundwork for deeper ecological validity studies that move towards more generalizable 6DoF movement in XR environments.

\section{Conclusion}

\par This paper examined the ecological validity of six-degree-of-freedom (6DoF) gestures for music mixing tasks in extended reality. Building on a design workshop, we evaluated how gestures for \textit{gain}, \textit{reverb}, and \textit{compressor} were perceived in terms of preference, cognitive load, and user experience.

\par Our results show trends that suggest that gesture preferences diverge by expertise and comfort, while cognitive load may remains stable across tasks. User experience varied by gesture type, highlighting tradeoffs between expressiveness and physical effort.

\par Together, these findings suggest that ecological validity in XR music mixing is shaped by embodied comfort and prior workflow familiarity. We hope this work informs the design of XR audio tools that balance expressiveness, learnability, and real-world mixing practices.


\bibliographystyle{ACM-Reference-Format}
\bibliography{references}

\appendix









\end{document}